\documentclass[12pt,a4paper]{article}
\usepackage[round, authoryear]{natbib}
\usepackage{amsmath}
\usepackage{amsbsy}
\usepackage{amscd}
\usepackage{graphicx}
\usepackage{afterpage}
\usepackage{longtable}
\usepackage{supertabular}
\usepackage{graphicx}
\usepackage{xcolor}

\begin{document}

\title{A flexible bivariate location-scale finite mixture approach to economic growth}
\author{Alessandra Marcelletti\footnote{%
Corresponding Author. Dipartimento di Economia Diritto ed Istituzioni,
Universit\`{a} di Roma ``Tor Vergata" via Columbia 2 00133 (RM) Italy,
e-mail: alessandra.marcelletti@uniroma2.com tel:+39-0672595649"}, Antonello Maruotti\footnote{Southampton Statistical Sciences Research Institute, University of Southampton} \footnote{Dipartimento di Scienze Politiche, Universit\`{a} di Roma Tre} and Giovanni Trovato\footnote{Dipartimento di Economia Diritto ed Istituzioni,
Universit\`{a} di Roma ``Tor Vergata"}}
\date{}
\maketitle
\begin{abstract}
We introduce a multivariate multidimensional mixed-effects regression model in a finite mixture framework. We relax the usual unidimensionality assumption on the random effects multivariate distribution. Thus, we introduce a multidimensional multivariate discrete distribution for the random terms, with a possibly different number of support points in each univariate profile, allowing for a full association structure. Our approach is motivated by the analysis of economic growth. Accordingly, we define an extended version of the augmented Solow model. Indeed, we allow all model parameters, and not only the mean, to vary according to a regression model. Moreover, we argue that countries do not follow the same growth process, and that a mixture-based approach can provide a natural framework for the detection of similar growth patterns. Our empirical findings provide evidence of heterogenous behaviors and suggest the need of a flexible approach to properly reflect the heterogeneity in the data. We further test the behavior of the proposed approach via a simulation study, considering several factors such as the number of observed units, times and levels of heterogeneity in the data.
\end{abstract}
\section{Introduction}
In modelling panel economic data, it is common to account for the unobserved heterogeneity between sample units, that is, the heterogeneity that cannot be explained by means of observable covariates (see e.g. \citealp{wooldridge2002} ; \citealp{fitzmaurice2008}). This is normally accomplished by the introduction of latent variables or random effects. For instance, a typical approach consists of associating a random intercept to every sample unit which affects the distribution of each time-specific response in the same fashion. This allows us to account for a form of unobserved heterogeneity which is due to unobservable covariates and related factors. The above considerations are obviously pertinent when we deal with economic growth modelling, where sample units (i.e. countries) are characterized by heterogeneous income performances. Addressing the heterogeneity of analyzed processes is of fundamental importance to the study to the economic growth and has led to a substantial evidence for the existence of variations in growth patterns across countries. Indeed, since Solow's seminal paper (1956), different econometric and statistical approaches are used to look at countries' growth. Dynamic panel data with fixed effect \citep{caselli1996, islam1995, temple1999}, as well as extreme bound analysis \citep{levine1992, temple2000}, Bayesian model averaging \citep{doppelhofer2000, fernandez2001} or model on varying coefficients are performed to deal with the main empirical challenges in growth theory: unobserved heterogeneity \citep{caselli1996, pesaran1995, lee1997, durlauf1995}, uncertainty \citep{temple2000} and omitted variable bias \citep{durlauf1999}.

Recently, data-driven approaches to estimate multiple (heterogeneous) growth processes have been employed within the wide class of mixture models (\citealp{alfo2008}, \citealp{owen2009}; \citealp{kerekes2012}; \citealp{bacsturk2012}; \citealp{bertarelli2013}). 

We propose an approach to panel growth data based on a flexible bivariate location-scale finite mixture approach, which may be seen as an extension of the approach introduced by \citet{alfo2008}. We introduce a bivariate bidimensional discrete random effects model to account for dependence between outcomes (i.e. per capita income and growth) and heterogeneity between countries in the augmented Solow growth model. The proposed approach may be cast in the literature about finite mixture models for panel data. It is worth noting that other extensions of the finite mixture approach for panel data are available in the literature. We mention, in particular, the extensions proposed by \citet{pittau2010} and \citet{martinez2011}, where countries are clustered into {\it clubs} depending on unobserved characteristics. Moreover, our approach is more general than those of \citet{durlauf1995} and \citet{ardicc2006} in which clustering is performed beforehand (i.e. clustering is exogenously specified). Indeed, we develop an endogenous clustering approach lying on a bivariate bidimensional model recovering \citet{bernanke2002} intuition: country's rate of investment and of human capital and the population growth rate are correlated with its long run growth of output per capita. Thus we contribute to this branch of literature by providing an empirical formulation of the augmented Solow model based on a multivariate-multidimensional specification, that allows to solve the unobserved heterogeneity issue. We address the heterogeneity issues related to: varying parameters across countries, omitted variables and non-linearities in the production function. Indeed, the incorrect specification of the country-specific effects leads to inconsistent parameter estimation, generating omitted variable bias \citep{caselli1996}.

As a by-product, we provide a posterior classification of countries sharing the same latent structure, highlighting strong heterogeneous behaviours. With respect to the existing approaches, we relax the assumption of the same posterior classification for the gross domestic product (GDP) per capita level and the growth rate. This allows us to let free the posterior classification given the observed variable and the latent effect, and to analyze the uncertainty and the variation in the different economics performance. We are able to distinguish between \textit{between} group, and \textit{within} group variations allowing for the human and physical capital and the population growth rate to simultaneously affect the different country growth experience, in terms of growth path and variability in the GDP per capita and growth rate. We further allow for explicitly modelling the scale parameter as a function of covariates. Indeed, we introduce two separates equations for the location and scale parameters of the dependent variables, such that the explanatory variables are associated not only to high or low values of the dependent variable, but also to the unpredictability of the variable itself.  

Computational complexity is often the price we have to pay to flexibility. However, we show that parameter estimates can be obtained by extending the Expectation-Maximization (EM) algorithm \citep{dempster1977} for finite mixture to the multidimensional case. Furthermore, we avoid any restriction on the covariance structure of the random effects as assumed e.g. by the so-called one-factor model \citep{winkelmann2000}, which is more parsimonious but could be hard to justify in empirical applications.
By allowing the number of mixture components to grow with the sample size, the proposed model can be also used as a semiparametric estimator of multivariate mixed effects models, where the distribution of the random effects is estimated by a discrete multivariate random variable with a finite number of support points. This can be seen as a possible solution to computational issues arising with multivariate mixed models. 

We illustrate the proposal by a  simulation study in order to investigate the empirical behaviour of the proposed approach with respect to several factors, such as the number of observed units and times and the distribution of the random term (with varying number of support points). Finally, we test the proposal by analysing a sample taken from the Summers-Heston
Penn World Tables (PWT) version 8.0 for years
1975-2005 for non-oil countries. We identify a set of variables that affect the volatility of economic growth and remark the importance of including {\it baseline} GDP as a covariate in the model specification. Moreover, different levels of heterogeneity are detected in GDP and GDP growth, respectively. More precisely, we find that our sample is much more heterogeneous with respect to GDP levels than growth patterns. Although this result sounds obvious, previous empirical results, based on unidimensional specification of the latent structure, were not able to distinguish for different heterogeneity levels \citep[see e.g.][]{alfo2008}. Instead, our approach can easily accommodate for different heterogeneity levels in the univariate profiles and, simultaneously, accounts for association between outcomes.  About obtained results, we get two clusters representing high-growth and low-growth countries, and six clusters are identified with respect to GDP levels.

The plan of the paper is as follows. In Section 2, we specify the proposed model in a general form and in Section 3 we provide
the computational aspects of the adopted maximum likelihood algorithm. In Section 4, we give a
comparison of the performance of several model specifications under different data generation schemes
by means of a simulation study. In Section 5, we present an empirical application on real world data motivating this paper. In
Section 6, we point out some remarks, along with drawbacks that may arise by adopting the proposed
methodology.

\section{Statistical framework}\label{stat}
We start assuming that the analysed sample is composed of $n$ statistical units (e.g. countries): continuous responses $y_{itj}$, corresponding to $(j=1, \dots, J)$ outcomes and two vectors of covariates ${\bf x}_{itj}' = (1,x_{itj1},\dots,x_{itjP_j})$ and ${\bf z}_{itj}' = (1,z_{itj1},\dots,z_{itjQ_j})$, which can vary over outcomes, are recorded for each unit $i$ ($i = 1,2,\dots,n$) at time $t$ ($t = 1,2,\dots,T$). Following the usual notation for longitudinal multivariate data, let ${\bf y}_{it} = (y_{i11},\dots,y_{itJ})'$ denote the vector of observed responses for unit $i$ at the $t$-th time. We assume that $y_{itj}$ are realizations of conditionally independent random variables, with parameters $\boldsymbol{\theta}_{itj}=(\theta_{itj1},\theta_{itj2},\dots,\theta_{itjM})$.
When we face multivariate analysis, and the primary focus of the analysis is not only to build a regression
model, but even to describe association among responses, the univariate approach is no longer sufficient and needs to be extended. In this context,
we are likely to face complex phenomena which can be characterized by having a non-trivial
correlation structure. For instance, omitted covariates may affect more than one response; hence,
modelling the association among the outcomes can be a fundamental aspect of research. Beyond
that, the association structure could be of interest by itself, as we may be interested in understanding
the nature of the stochastic dependence among the analysed phenomena. Furthermore, it is well
known that, when responses are correlated, the univariate approach is less efficient than the
multivariate one, since in estimating the parameters in the single equations, the multivariate
approach takes into account of zero restrictions on parameters occurring in other equations (for a detailed discussion on this topic see e.g. \citealp{zellner1962}; \citealp{davidson1993}).

A standard way to insert dependence among responses is to assume that they share some common latent structure. Thus, the model specification is completed by connecting the $J$ univariate submodels through a common latent structure, represented by a set of random effects ${\bf u}_i = ({\bf u}_{i1},\dots,{\bf u}_{iJ})$ which account for potential heterogeneity among statistical units and correlation between outcomes. In a regression setting, the interest is usually focused upon the mean which is modelled through a linear mixed model, providing a very broad framework for modelling dependence in the data \citep{verbeke2014}. Nevertheless, statistical models rarely allow the modelling of parameters other than the
mean of the response variable as functions of the explanatory variables. For instance, the scale parameter is usually not modelled explicitly in terms of the explanatory variables but implicitly through its dependence on the mean. In the following, we relax such a constrain and define a location-scale multivariate regression framework by specifying $J$ conditionally independent (given the covariates and the random effects) regression models. Let us decompose the design vector as ${\bf x}_{itj} = \{{\bf x}_{itj}^{(1)},{\bf x}_{itj}^{(2)}\}$, where the variables whose effects are assumed to be fixed are collected in ${\bf x}_{itj}^{(1)}$, while those which vary across units are in ${\bf x}_{itj}^{(2)}$. The $M$-dimensional parameter vector $\boldsymbol{\theta}_{itj}$ is related to covariates and random effects. Let us specify $\theta_{itj1}$ as the location parameter,  $\theta_{itj2}$ as the scale parameter and  $\theta_{itj3}$ as a shape parameter (whenever needed) and let $g_m(\cdot)$ be a known monotonic link function relating  $\theta_{itjm}, m = 1,\dots,3$ to covariates and random effects, we define the following regression models

\begin{equation}
\left\{%
\begin{array}{l}
g_1(\theta_{itj1})={\bf x}_{itj}^{'(1)}\boldsymbol{\lambda}_j+{\bf x}_{itj}^{'(2)}{\bf u}_{ij}\\
\\
g_2(\theta_{itj2})={\bf z}_{itj}'\boldsymbol{\gamma}_j
\\
\\
g_3(\theta_{itj3})=\boldsymbol{\tilde{\gamma}}_j
\end{array}
\right. \label{model}
\end{equation}

where ${\bf u}_{ij}$ represents unit- and outcome-specific random effects, drawn from a multivariate parametric density, $\boldsymbol{\lambda}_j$, $\boldsymbol{\gamma}_j$ and $\boldsymbol{\tilde{\gamma}}_j$ are outcome- and moment-specific fixed parameters. Of course, covariates may be included in the shape-parameter model, but this may complicate results interpretation in empirical applications.

 Given the model assumptions, the likelihood function can be written as follows:
\begin{equation}
L(\cdot)= \prod\limits_{i=1}^{n} \left\{\int_{\mathcal{U}}\prod\limits_{j=1}^{J}%
\prod_{t=1}^{T}f(y_{itj}\mid {\bf u}_{ij}, {\bf x}_{itj}, {\bf z}_{itj})b({\bf u}_{i})d{\bf u}_{i}\right\}
\end{equation}
where $f(\cdot)$ is a generic probability density function, $\mathcal{U}$ represents the support for $b({\bf u}_i)$, the distribution function of ${\bf u}_i$, with $E({\bf u}_i)=0$.

Although, at first glance, the approach proposed so far is appealing, it has several computational drawbacks and limitations. Indeed, the random effects distribution is unknown and assuming a multivariate Gaussian distribution may be a too strong and unverifiable assumption and, moreover, may affect parameters estimate. Indeed, in some situations, the distribution of the random effects may depart from normality. This problem has been addressed, for example, by specifying a different parametric distribution family for the random terms, such as multivariate skewed and/or heavy-tailed distributions (\citealp{ferreira2006}; \citep{ferreira2004}). An alternative approach is to use nonparametric maximum likelihood based on finite mixtures, which provide a more flexible framework to deal with departure from normality of the random effects distribution (see e.g. \citealp{bohning1995}; \citealp{aitkin1999}). Nevertheless, even if the latter is computationally efficient when compared to parametric random effect models, it is intrinsically unidimensional, since
it is based on a single categorical latent variable. This may lead to problems
when the task is testing for dependence between the random effects. Indeed, the model under independence
does not occur as a special case of the dependence model.\\
In the following, we consider a $J$-variate $J$-dimensional latent structure such that the independence model is nested in the multivariate one, and different levels of heteorgeneity in the $J$ univariate profiles can be identified. In order to specify a latent structure of this kind, we leave the distribution of the random effect $b(\cdot)$ completely unspecified and invoke the non-parametric maximum likelihood approach.

Formally, random effects distribution can be approximated through a discrete distribution with $K_j \leq n$ support
points at the marginal level. Mass joint probability $\pi_{k_{1},k_{2},\dots,k_J}$ are attached to location $({\bf u}_{k_{1}},{\bf u}_{k_{2}},\dots,{\bf u}_{k_{J}})$ for $k_j = 1, \dots, K_j $. Focusing on the bivariate ($J=2$) case, without lacking of generality, we define the following location-scale multivariate regression model
  
\begin{equation}
\left\{%
\begin{array}{l}
g_1(\theta_{itj1})={\bf x}_{itj}^{'(1)}\boldsymbol{\lambda}_j+{\bf x}_{itj}^{'(2)}{\bf u}_{k_j}\\
\\
g_2(\theta_{itj2})={\bf z}_{itj}'\boldsymbol{\gamma}_j
\\
\\
g_3(\theta_{itj3})=\boldsymbol{\tilde{\gamma}}_j
\end{array}
\right. \label{model_np}
\end{equation}

According to model assumptions, the likelihood function in the bivariate case is given by
\begin{equation}
L(\cdot)=\prod\limits^{n}_{i=1}\left\lbrace\sum\limits^{K_{1}}_{k_{1}=1}\sum%
\limits^{K_{2}}_{k_{2}=1}\pi_{k_{1}k_{2}}\prod\limits_{j=1}^{2}\prod%
\limits^{T}_{t=1}f(y_{itj}|{\bf x}_{itj},{\bf z}_{itj},{\bf u}_{i1}={\bf u}_{k_1},{\bf u}_{i2}={\bf u}_{k_{2}})\right\rbrace \label{lik}
\end{equation}

where $\pi_{k_{1}k_{2}}=Pr({\bf u}_{i1}={\bf u}_{k_1},{\bf u}_{i2}={\bf u}_{k_{2}})$ is the joint probability associated to each couple of locations $({\bf u}_{k_1},{\bf u}_{k_2})$. The following constraints hold $\sum\limits^{K_{1}}_{k_{1}=1}\pi_{k_{1}}=\sum\limits^{K_{2}}_{k_{2}}%
\pi_{k_{2}}=\sum\limits_{k_{1}k_{2}}\pi_{k_{1}k_{2}}=1$

with 

$$\pi_{k_{1}}=Pr({\bf u}_{i1}={\bf u}_{k_{1}})=\sum_{k_{2}=1}^{K_2}\pi_{k_{1}k_{2}}$$
 and $$\pi_{k_{2}}=Pr({\bf u}_{i2}={\bf u}_{k_{2}})=\sum_{k_{1}=1}^{K_1}\pi_{k_{1}k_{2}}.$$
We would remark that the number of locations (i.e. mixture components) may vary between outcomes. Thus, we control for heterogeneity in the univariate
profiles and for the association between latent
effects in the two profiles. This approach results in a finite mixture with $K_1\times K_2$ components, in which each of the $K_1$ locations are coupled with each of the $K_2$ locations of the second outcome. If $J=1$, our proposal reduces to a univariate finite mixture model. 
%
%


\section{Computational details} 
Let  $\boldsymbol{\tilde{\theta}}$ be a short-hand notation for all non-redundant models parameters corresponding to the vectors $(\mbox{\boldmath$\lambda $}, \mbox{\boldmath$\gamma $}, \mbox{\boldmath$\tilde{\gamma} $},\mbox{\boldmath$\pi$}, \mbox{\boldmath$u $})$, inference for the proposed model is based on log-transformation of the likelihood in (\ref{lik}).

To estimate $\boldsymbol{\tilde{\theta}}$, we maximized the log-transformation of (\ref{lik}) by using a version of the EM algorithm (Dempster et al., 1977). The EM algorithm alternates the following steps until convergence
\begin{description}
\item[E-step:] compute the conditional expected value of the complete data log-likelihood given the observed data and the current estimate of model parameters; and
\item[M-step:] maximize the preceding expected value with respect to $\boldsymbol{\tilde{\theta}}$.
\end{description}

Let $w_{ik_1k_2}$ denote a dummy variable equal to 1 if unit $i$ is in component $k_1$ and $k_2$ in the two univariate profiles, respectively, and zero otherwise. The complete data likelihood, which we would compute if we knew these dummy variables, is 

\begin{equation}
L_c(\cdot)=\prod_{i=1}^n\left[ \sum\limits_{k_{1}=1}^{K_1} \sum\limits_{k_{2}=1}^{K_2} \pi_{k_{1}k_{2}} f_{ik_{1}k_{2}}\right] ^{w_{i k_{1}k_{2}}} \label{compLK}
\end{equation}

And its corresponding log-transformation is
\begin{equation}
\ell_c(\cdot)=\sum\limits_{i=1}^n\sum\limits_{k_{1}}^{K_1} \sum\limits_{k_{2}}^{K_2} w_{k_{1}k_{2}}\left\lbrace \log (\pi_{k_{1} k_{2}})+\log f_{ik_{1}k_{2}}\right\rbrace  \label{complogLK}
\end{equation}
where $f_{ik_{1}k_{2}}=f_{ik_{1}}f_{ik_{2}}=\prod_{t=1}^Tf(y_{it1}|{\bf x}_{it1}, {\bf z}_{it1}, u_{k_{1}}) f(y_{it2}\mid {\bf x}_{it2}, {\bf z}_{it2}, u_{k_{2}})$.

The conditional expected value of $\ell_c(\cdot)$ at the E-step has then the same expression as given previously in which we substitute the variable $w_{ik_1k_2}$ with its corresponding expected value
\begin{equation}
\hat{w}_{k_{1}k_{2}}=\frac{\pi_{k_{1}k_{2}} f_{ik_{1}k_{2}} }{\sum\limits_{k_{1}k_{2}}\pi_{k_{1}k_{2}}f_{ik_{1}k_{2}}}.\label{post}
\end{equation}

\noindent where $\hat{w}_{k_{1}k_{2}}$ is the posterior probability the the $i$-th unit belongs jointly to the $k_{1}$ and $k_{2}$ components of the mixture. We can easily get the marginal posterior probabilities
\begin{equation}
\hat{w}_{ik_{1}}=\sum\limits_{k_{2}}\hat{w}_{ik_{1}k_{2}} \quad \hat{w}_{ik_{2}}=\sum\limits_{k_{1}}\hat{w}_{ik_{1}k_{2}}
\end{equation} 

At the M-step, the conditional expected value of (\ref{complogLK}) is maximized by separately maximizing its components. Indeed, the score function is 
$$\sum\limits_{i=1}^n\sum\limits_{k_{1}}^{K_1} \sum\limits_{k_{2}}^{K_2} w_{k_{1}k_{2}}\frac{\partial}{\partial\boldsymbol{\theta}}\left\{ \log (\pi_{k_{1} k_{2}})+\log f_{ik_{1}} + \log f_{ik_{2}}\right\}.$$ 

Let us partition the parameter vector $\boldsymbol{\tilde{\theta}} = (\boldsymbol{\tilde{\theta}}_{k_1},\boldsymbol{\tilde{\theta}}_{k_2})$, where $\boldsymbol{\tilde{\theta}}_{k_j}$ collects the parameters of the $j$-th profile such that

\begin{equation}
\frac{\partial \ell(\cdot)}{\partial  \mbox{\boldmath$\tilde{\theta} $}_{k_{1}}}= \sum_{i=1}^n\hat{w}_{ik_1}\frac{\partial}{\partial\boldsymbol{\tilde{\theta}}_{k_1}}\log(f_{ik_1})\label{M1};
\end{equation}
\begin{equation}
\frac{\partial \ell(\cdot)}{\partial  \mbox{\boldmath$\tilde{\theta} $}_{k_{2}}}= \sum_{i=1}^n\hat{w}_{ik_2}\frac{\partial}{\partial\boldsymbol{\tilde{\theta}}_{k_2}}\log(f_{ik_2})\label{M2}
\end{equation}

and 
\begin{equation}
\frac{\partial\ell(\cdot)}{\partial {\pi}_{k_{1}k_{2}}}=\sum\limits_{i=1}^n\hat{w}_{ik_{1}k_{2}}\frac{\partial}{\partial \pi_{k_{1}k_{2}}} \log \pi_{k_{1}k_{2}}\label{Mpi}
\end{equation}

An explicit solution is available to maximize the last M-step equation, which consists of $$\hat{\pi}_{k_1k_2} = \frac{\sum_{i=1}^n\hat{w}_{ik_1k_2}}{n}.$$ To maximize the other two parts, we can use a standard iterative algorithm of Newton-Raphson type for linear mixed models. We take the value of $\boldsymbol{\tilde{\theta}}$ at convergence of the EM algorithm
as the maximum likelihood estimate. As it is typical for finite mixture models the likelihood may be multimodal and the point at convergence depends on the starting values for the parameters, which then need to be carefully chosen. In this regard, we run the EM algorithm from multiple random starting points for a number of steps, then pick the one with the highest likelihood, and continue the EM from the picked point until convergence. However, other methods can be used; for example, a gradient function based on directional derivatives can be used to get optimality criteria (see e.g. \citealp{wang2010}). 

At last, we approach the model selection problem by looking at penalized likelihood criteria, Akaike information criterion (AIC) and Bayesian information criterion (BIC). In this way we select the number of mixture components and we can also compare the different models. BIC, achieved in the Bayesian framework is found to be satisfactory in the model-based clustering context (see among others \citealp{fraley2002}, for further details). Both criteria are likelihood based and they differ for the different penalization used. In fact, denoting with $d$ the number of independent parameters to be estimated and with $n$ the sample size, BIC is obtained as $BIC=-2\ell (.) + d \ln(n)$, and AIC is given by $AIC=-2\ell (.) + 2*d$.


\section{Simulation study}
To assess the properties of the maximum likelihood estimator described in Section 3, we carried out a simulation study, which is described subsequently. The same study allows us to assess the goodness of classification.

\subsection{Simulation design}
We considered two scenarios: the first with two response variables (both Gaussian-distributed) with $K_1=K_2=2$ mixture components each and the second with higher heterogeneity levels, i.e. by defining a bivariate model with $K_1=2$ and $K_2=3$ mixture components for each outcome respectively. Under each scenario, we considered two continuous covariates, one in the linear predictor for the mean and one in the regression model for the scale parameter, and generated 500 samples from the proposed model with $T = 5; 10$ (panel length) and $n = 100; 1000$ (sample size). Under this setting, $\boldsymbol{\theta}_{itj} = (\theta_{itj1},\theta_{itj2})= (\mu_{itj},\sigma_{itj})$

{\it Scenario 1}. We assume that the
outcomes are conditionally independent and proceeded to generate 500 samples from $$Y_{it1}\mid \mu_{it1},\sigma_{it1} \sim N(\mu_{it1},\sigma_{it1})$$ $$Y_{it2}\mid \mu_{it2},\sigma_{it2} \sim N(\mu_{it2},\sigma_{it2})$$
where the following bivariate regression model (with a single covariate) holds
$$\mu_{it1} = u_{k_1}+\lambda_{11}x_{it}=\left\{\begin{array}{cc}-1 + 0.5x_{it},&k_1=1\\1 + 0.5x_{it},&k_1=2\end{array}\right.$$
$$\log(\sigma_{it1}) =\gamma_{01}+\gamma_{11}z_{it}= 0.5+0.75z_{it}$$
and
$$\mu_{it2} = u_{k_2}+\lambda_{12}x_{it}=\left\{\begin{array}{cc}2 + 0.5x_{it},&k_2=1\\-2 + 0.5x_{it},&k_2=2\end{array}\right.$$
$$\log(\sigma_{it2}) =\gamma_{02}+\gamma_{12}z_{it}= 1+0.25z_{it}$$

with $$\boldsymbol{\pi} = \left[\begin{array}{cc}\pi_{11}&\pi_{12}\\\pi_{21} & \pi_{22}\end{array}\right] = \left[\begin{array}{cc}0.4&0.1\\0.2 & 0.3\end{array}\right]$$

{\it Scenario 2}. We assume that the
outcomes are conditionally independent and proceeded to generate 500 samples from $$Y_{it1}\mid \mu_{it1},\sigma_{it1} \sim N(\mu_{it1},\sigma_{it1})$$ $$Y_{it2}\mid \mu_{it2},\sigma_{it2} \sim N(\mu_{it2},\sigma_{it2})$$
where the following bivariate regression model (with a single covariate) holds
$$\mu_{it1} = u_{k_1}+\lambda_{11}x_{it}= \left\{\begin{array}{cc}-1 + 0.5x_{it},&k_1=1\\1 + 0.5x_{it},&k_1=2\end{array}\right.$$
$$\log(\sigma_{it1}) =\gamma_{01}+\gamma_{11}z_{it}= 0.5+0.75z_{it}$$
and
$$\mu_{it2} =  u_{k_2}+\lambda_{11}x_{it}=\left\{\begin{array}{cc}2 + 0.5x_{it},&k_2=1\\-2 + 0.5x_{it},&k_2=2\\ 0 + 0.5x_{it},& k_2 = 3\end{array}\right.$$
$$\log(\sigma_{it2}) = \gamma_{02}+\gamma_{12}z_{it}=1+0.25z_{it}$$

with $$\boldsymbol{\pi} = \left[\begin{array}{ccc}\pi_{11}&\pi_{12} & \pi_{13}\\\pi_{21} & \pi_{22} & \pi_{23}\end{array}\right] = \left[\begin{array}{ccc}0.1&0.1 & 0.2\\0.2 & 0.3 &0.1\end{array}\right].$$

\subsection{Simulation results}
For each sample, we computed the maximum likelihood estimate of the parameters and the corresponding standard errors, under the assumed model. We also evaluate the performance of the proposed in correctly clustering the statistical units into mixture components. The Rand Index \citep{hubert1985} is considered. The true matrix ${\bf W} = \{w_{ik_1k_2}\}$ of component membership and the crispy estimated matrix ${\bf W}^* = \{w_{ik_1k_2}^*\}$, where each element $w_{uk_1k_2}^*$ is defines as $$w_{uk_1k_2}^*=\left\{\begin{array}{cc}1 & {\rm if} k_1,k_2 = \arg\max_{k_1,k_2}\hat{w}_{ik_1k_2}\\ 0 & {\rm otherwise}\end{array}\right.$$ are compared. Formally, let $n_{k_1k_2}$ denote the number of all pairs of data points which are either put into the same cluster by both partitions or put into different clusters by both partitions. Conversely, let $n^*_{k_1k_2}$ denote the number of all pairs of data points that are put into one cluster in one partition, but into different clusters by the other partition. The partitions disagree for all pairs $n^*_{k_1k_2}$ and agree for all pairs $n_{k_1k_2}$. We can measure the agreement by the Rand index $n_{k_1k_2}/(n_{k_1k_2}+n_{k_1k_2}^*)$ which is invariant with respect to permutations of cluster labels.

For {\it Scenario 1}, the simulation results in terms of bias and standard deviation of the maximum likelihood estimator of each parameter of interest are shown in Table \ref{t:sc1}, together with the Rand Index. We can observe that, the bias of each estimator is always low and decreases as $T$ increase; moreover, its standard deviation decreases. Indeed, for $n=100$  and $T=10$ the estimators are unbiased. By increasing the number of available times, the clustering performance improves as well as shown by the Rand Index. For sake of brevity, we do not report the results for $n=1000$. They do not provide any further insight to the already discussed results.

By considering {\it Scenario 2}, in which a higher degree of heterogeneity is assumed in one of the two outcomes, we can easily detect a different estimators behavior (see Table \ref{t:sc2}). Obviously, for small sample size ($n=100$) and $T=5$, higher bias and standard deviations are estimated with respect to those in {\it Scenario 1}. However, estimates variability decreases at the expected rate of $\sqrt{n}$ with respect to $n$ and at a faster rate with respect to $T$. By increasing the sample size to $n=1000$, we get less biased estimates, as expected. Clustering performances are sensitive to $n$ and $T$ as well. Indeed ,the larger is the sample size the better is the recovered latent structure.

\section{Empirical framework}
\subsection{Data}

The sample is composed by an unbalanced panel of 101 countries over the period 1975-2010. Data on the dependent variables and the investment share on physical capital \textit{(sk) }are retrieved from the Heston-Summers-Aten dataset (Penn World Table 8.0). Data on human capital \textit{(sk)}, measured as the total enrollment in secondary education, is retrieved from the World Bank. From the same database, we also collect: openness to trade \textit{(open)}, measured as the sum of exports and imports as share of GDP,  and the credit to the Private Sector as a fraction of GDP \textit{(fin)}, used as a proxy for financial development. In order to understand the effect of financial factor on the growth fluctuations through the household consumption channel, the private sector on GDP is preferred as measure since it does not account for the credit provided from the Central and development bank to the public sector.  
Government consumption \textit{(govcons)} is calculated as the general government final consumption expenditure (as share of GDP). Unemployment rate\textit{(unempl)} and the inflation level (\textit{infl}) are obtained from the Penn World Table 8.0 dataset. 

In order to avoid the endogeneity problems related to growth model estimation, we consider non-overlapping 5-year period with explanatory variable averaged over the corresponding time period; while the dependent variables are taken 5 periods ahead \citep{bond2001}. Indeed, endogeneity could be due to the fact that ``country-specific heterogeneity
cannot be captured if one does not look at between-countries variation which cannot be explained
by observed covariates but remains persistent over the analysed time period." (\citealp[pg. 495]{alfo2008}). Thus, the dependent variables are the average of GDP per capita over the 5-years period ($y_{it1}$), and the average annual growth of real GDP over the same non overlapping period ($y_{it2}$). Table \ref{t:sum} provides descriptive statistics, variables description, and data sources. 

\begin{figure}\centering \caption{Histograms of response varialbes}\label{hist}
\includegraphics[scale=0.36]{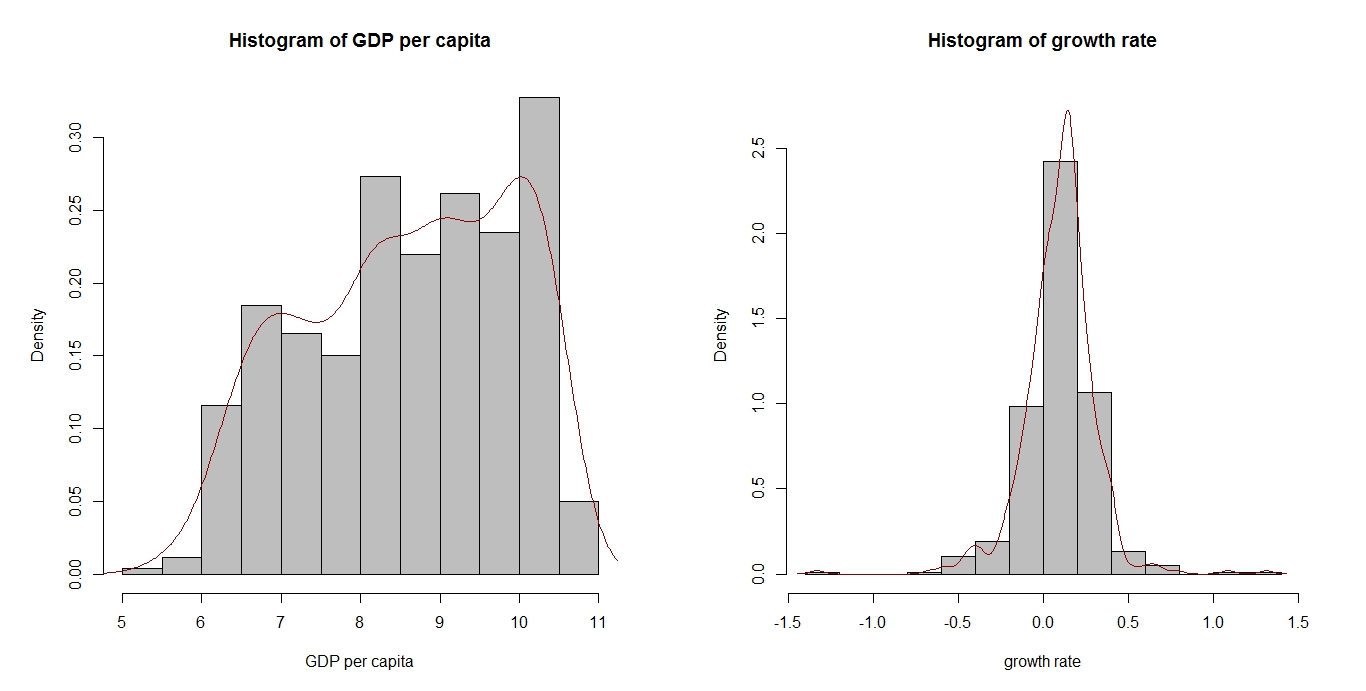}
\end{figure}

To analyze the marginal distribution of the response variables, graphical and statistical analysis are provided. Figure \ref{hist} displays a clear multimodal distribution for the GDP level, supporting the idea of different sub-populations in the outcome. The marginal distribution of growth rates does not show any multimodality, although a small bump can be detected on the left with respect to the distribution mode. However, we cast some doubts that growth rate follows a Gaussian distribution. Thus, to complement the graphical analysis, Shapiro-Wilk and Jarque-Bera tests and summary statistics are provided in Table \ref{sumstatres} for the two outcomes. Skewness and kurtosis of each response variable indicate a departure from the normal distribution. Whilst, it is expected that both Shapiro-Wilk and Jarque-Bera tests indicate departure from marginal normality for the GDP level, we obtain a significant departure from normality for the growth rate outcome as well. Thus, we opt for a (mixture of) heavy-tailed distribution to properly model growth rates.


\subsection{Economic growth}\label{growth}
 
To understand the cross-country differences in income performances and to account for dependence between per capita income and growth, we introduce a flexible bivariate multidimensional finite mixture approach for the location and the scale parameters, and for the shape parameter when it is required, as described in Section 2. 
To jointly determine the evolution of income per capita and volatility of growth, instead of modelling the scale parameter through the dependence on the mean, we explicit the variance of the growth rate as dependent on explanatory variables. Thus, growth determinants are associated not only to high or low values of the dependent variable but also to unpredictability of the variable itself.


Formally, for each country $i$ at time $t$, let the GDP level ($y_{it1}$) be a Gaussian random variable, i.e. $y_{it1}\sim N(\mu_{it1},\sigma_{it1})$, and the GDP growth rate ($y_{it2}$) be t-distributed to account for heavy tails in the growth distribution, i.e. $y_{it2}\sim t(\mu_{it2},\sigma_{it2},\nu_{it2})$ . To explore the determinants of both growth level and growth volatility, we choose variables found to be robust in the economic growth literature (see e.g. \citealp{levine1992, mankiw1992, cecchetti2006}), and define the following mixed-effects regression model for $y_{it1}$

\begin{equation}
\left\{%
\begin{array}{l}
\mu_{it1}=u_{i10}+ \lambda_{11} sk_{it} + \lambda_{21} sh_{it} + \lambda_{31} (n_{it}g\delta)\\
\\
\log(\sigma_{it1})={\gamma}_{01}
\\

\end{array}
\right. \label{modelgdp}
\end{equation}

\noindent where $sk_{it}$ and $sh_{it}$ are the share of output invested in
physical and human capital, respectively, $\delta$ is the depreciation rate, $n$ is the population growth rate and $g$ is the technological progress. As it is common in the growth literature, the term $g+\delta$ is assumed to be common across countries and equal to 0.5. Parameters in model (\ref{modelgdp}) capture the effect of the human and physical capital accumulation process, and the population growth on the income per capita. They can be explicit as:
\begin{equation}
\lambda_{11}=\frac{\alpha}{(1-\alpha-\beta)} \quad \lambda_{21}=\frac{\beta}{(1-\alpha-\beta)} \quad \lambda_{31}=\frac{\alpha+\beta}{(1-\alpha-\beta)}
\end{equation}

\noindent where $\alpha$ and $\beta$ are respectively the share of physical and human capital, such that $(\alpha + \beta)<1$. It is worth noting that the $\lambda_{11}$ and $\lambda_{21}$ are expected to be positive, while $\lambda_{31}$ to be negative, since human and physical capital accumulation boost economic growth, while the population growth rate is thought to discourage the evolution of the economy (see among others \citealp{solow1956, mankiw1992, barro1991}).
The random intercept $u_{i10}$ is let free to vary across countries since it captures the unobserved heterogeneity due to the omission and/or the immeasurable nature of some country-specif factors. 

According to \citet{bernanke2002}, the definition of the augmented Solow model implies a bivariate growth model, in which the long run growth of output per capita is correlated with the accumulation of human and physical capital and the population growth rate.
We adopt a reduced-form model for the location parameter of the growth rate (see \citealp{goetz1996}for further details) such that

\begin{equation}
\left\{%
\begin{array}{l}
\mu_{it2}=u_{i20}+u_{i21}\ln(yc_{it})\\
\\
\log(\sigma_{it2})=\gamma_{0}+ \gamma_{12} unempl_{it} + \gamma_{22}fin_{it} + \gamma_{32}infl_{it} +\gamma_{42}open_{it} +\gamma_{52}govcons_{it} \\
\\
\nu_{it2}={\tilde{\gamma}}_{02}
\end{array}
\right. \label{model}
\end{equation}
The random coefficient $u_{i21}$ (attached to the initial level of income per capita) controls for the transitional dynamics affecting the evolution of the growth rate. It is worth recalling that the neoclassical approach predict a fixed and negative coefficient for the initial level of income per capita $lnyc_{it}$ accounting for country convergence. \\
In our approach, economic stability is directly modelled by including an equation for the variance of the growth rate, that regress the unpredicatability of the response variable on financial development, international openness, government consumption, inflation and unemployment rate (e.g., \citealp{cecchetti2006, giovanni2009}). We expect that cyclical variables (unemployment rate and inflation) have a destabilizing effect on growth, i.e. $\gamma_{12}$ and $\gamma_{32}$ are expected to be positive, while financial development and government consumption decrease growth volatility. The effect of openness to trade on economic growth is still debated in the literature.   \\
Again, the random terms $u_{i02}$ and $u_{i21}$ in the location parameter's equation are let free to vary among countries and response variables, by allowing for a random slope as well. This allows us to simultaneously understand the variation across country in the standard of living and in the volatility of the outcome per capita, leaving the posterior classification of the mixture model to be free to vary among outcomes.\\




\subsection{Results}
\begin{figure}\centering \caption{Model fitting: GDP level (left box), GDP growth (right box)}\label{f:fit}
\includegraphics[scale=0.36]{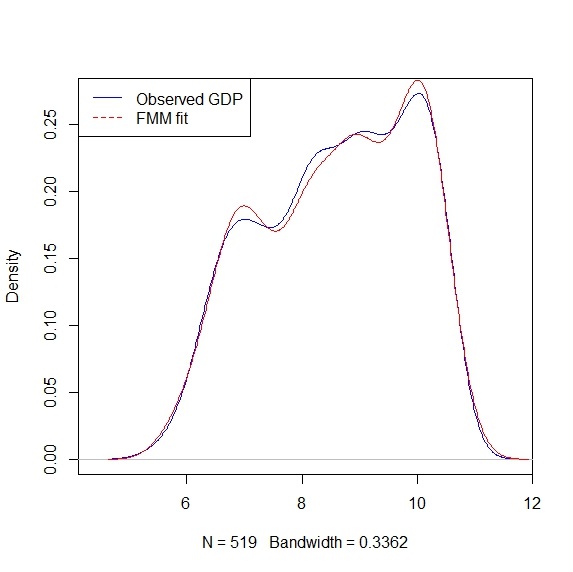}
\includegraphics[scale=0.36]{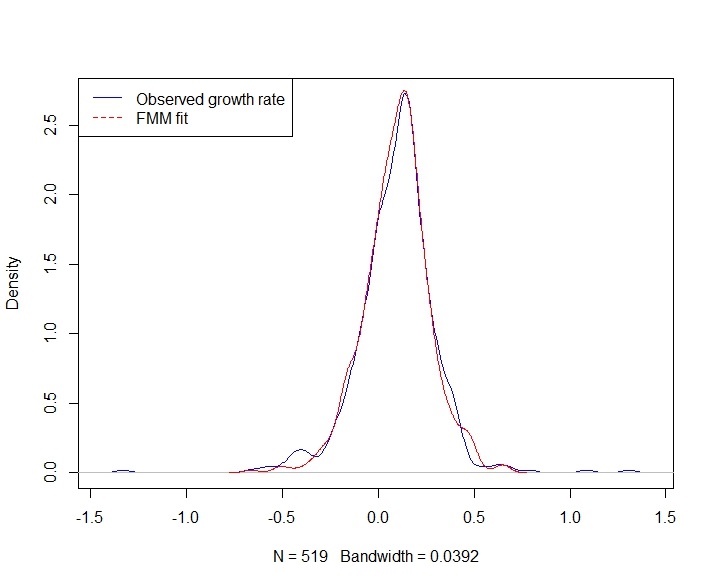}
\end{figure}

A major research question would concern the need of a complex model like the one we introduce to properly model economic growth. Thus, to remark the crucial role of the bivariate approach with respect to the univariate one, we start our empirical analysis by comparing univariate and multivariate approaches. Firstly, we fit univariate mixed-effects models for each outcome separately, with $K_1=2,\dots,7$ and $K_2=2,\dots,4$. Model selection results are provided in Table \ref{t:AICBICuniv}, and models with $K_1=6$ and $K_2=2$, respectively, are selected. Similarly, we perform model selection for the bivariate model specified in the previous section, with varying $K_1=2,\dots,7$ and $K_2=2,\dots,4$.  In the bivariate case the AIC is in favour of the $K_1=6$ and $K_2=3$, while the BIC select the model with $K_1=6$ and $K_2=2$ groups (see Table \ref{t:AICBICbiv}). By comparing penalized likelihood criteria, it is clear that linking the two univariate profiles by a shared (correlated) random effects structure, i.e. adopting a bivariate approach, leads to better results in terms of trade-off between model fit and model complexity. In the following we look at the results obtained with the bivariate selects according to the BIC. This choice is motivated by looking at parsimony and for comparison purposes (with respect to univariate model specifications). In Figure\ref{f:fit} we provide evidence of the goodness of fit of the proposed model, and of the relatively small increase in goodness of fit the $K_1=6$ and $K_2=3$ model selected according to the AIC. The Parameter estimates are provided in 
Table \ref{res}. The main difference between the univariate and the multivariate approaches is on the magnitude of covariates effects in the equation for the mean of GDP level. Indeed, the bivariate approach parameter estimates confirm the augmented Solow model intuition, i.e. the accumulation process of physical and human capital exhibits more reasonable value of the coefficients with respect to univariate case. 
As discussed before, the intercept term captures the omitted country-specific features, such as, above all, institutional characteristic. This is related to the idea that accumulation driven growth equation is incomplete (see e.g. \citealp{alfo2008}), and, coherently with the literature, the highest value for the random effect is found for the component clustering the richest and more industrialized countries, such as USA and UK. However, we will investigate the obtained clustering in depth in Section 5.4.

As formalized before, the location parameter for the growth rate is estimated by applying a reduced-form model where the independent variables is the 5-years backward value of GDP per capita. This allows for avoiding biased estimation in the parameters due to the dependence among physical and human capital on income per capita \citep{goetz1996}. Furthermore, to account for the difference in initial level of GDP per capita, we leave the initial level of GDP to vary among countries. Results show the existence of two groups: the first group characterized by a negative and significant effect of the initial level of GDP on the growth pattern, confirming economics theory about convergence; the second group is characterized by the possible existence of multipla equilibria and the lack of convergence. 
These results suggest the presence of a convergence club, that is, a group of countries with different levels of per capita real GDP within which countries converge to a group-specific growth path, i.e. the neoclassical prediction of the convergences is proved for those countries. The second component, clustering low income countries, shows lack of income convergence allowing for the potential existence of multipla equilibria, as obtained by \citet{owen2009}. To summarize, accounting for heterogeneity, we can conclude for the existence of two difference of groups in the growth process: one in which countries converge and one in which the positive and significant coefficient associated to the initial level of GDP per capita suggests the lack of convergence and the possible existence of multipla equilibria.

The volatility of growth rate is mainly due to the unemployment rate and to the financial development. This implies that changing in the labor market and in the financial sector are the main causes of the economics, respectively, instability and stability. The high level of financial development is found to be negatively related to the growth variability. This could be due to the direct connection between the financial development and the household consumption. As \citet{aghion1999}, and \citet{easterly2001} suggest, an increase in the private credit to GDP generates more consumption smoothness, by reducing the household liquidity constraints; in turn, the less consumption volatility (smoothed by the less liquidity constraints) leads to less volatility in growth. 
Unemployment is found here to play a destabilizing role on output fluctuation. This could be due to the fact that an increase in the unemployment level generates a decrease in consumption. Inflation, openness to trade and government consumption are found to be non significantly different from zero in the bivariate equation for the scale parameters (see Table \ref{res}). 

An high level of openness to trade is associated to an improvement in the financial and commercial risk sharing with foreign countries \citep{cecchetti2006} and to a consequent increase in the vulnerability to the demand and supply shock \citep{newbery1984}. On the other hand, stabilizing effect of the openness to trade could be due to the financial structure of country itself, i.e. the most exposed to capital flows, the most stabilizing effect on growth openness to trade \citep{cavallo2008}, or to the degree of diversification of exports \citep{haddad2013}. Furthermore, we obtain that cyclical fluctuations in the growth rate are negatively related to the labour market participation \citep{okun1962} and to the inflation rate. 

\subsection{Clustering}
An interesting by-product of our approach is the possibility to cluster 
countries on the basis of their posterior probabilities ${w}_{ik_ {1}k_{2}}
$. The i-th country can be classified in the $k_1 - k_2$-th group if $
\hat{w}_{ik_{1}k_{2}}=max_{k_{1}k_{2}} (\hat{w}_{i11}, \dots,  \hat{w}_{iK_{1}K_2})
$. It is worth nothing that each group is characterized by homogeneous values 
of (estimated) random effects; thus, conditionally on observed covariates, 
countries clustered in the same group share a similar behaviour with respect 
to the event of interest (i.e. GDP level and growth). This represents a 
substantial difference with conclusion derived by assuming any parametric 
approach for the random terms. \\
Table 8 displays the a posteriori classification. With respect to the GDP level groups, $k_1 = 1$ and $k_1 = 6$ cluster well-developed countries (with any few exceptions), while the poorest countries are clustered in $k_1=4$. It is interesting to notice that high levels of GDP are often associate to higher propensity to grow. Indeed, all countries (but Costa Rica, Mexico, Panama, Turkey and Venezuela) clustered in $k_1=1 $ or $k_1=6$ are assigned to $k_2=1$, i.e. the growth group with the highest propensity to growth, somehow alleviated by the initial GDP level. Similarly, the ``poorest countries" share a lower propensity of economic growth with the exception of China and Thailand (as expected).\\
The obtained classification is, in this case, not only a mathematical tool able to capture the unobserved heterogeneity, but groups may have a ``physical" meaning. Indeed, countries in the same cluster often share similar technological, institutional and/or geographical characteristics (e.g. OECD countries  are clustered together), and in general a similar socio-economic background. \\
A final remark concerns the impact of initial GDP level on growth because it it important to check for convergence. Our results suggest two different process. The first one involves developed countries, whose growth is relatively high and in which higher values of GDP contributes to the growth process, thus leading to ``convergence". On the other hand, for ``poorest" countries differences will increase as the initial GDP positively affects economic growth leading to divergence. 
\section{Conclusion}
In this paper we introduce a flexible multivariate multidimensional random model allowing for all model parameters to depend on covariates in a regression framework. We relax the common unidimensionality assumption of the random effects distribution, allowing for a general and flexible association structure among the outcomes. The proposed approach is motivated by the analysis of economic growth in presence of heterogeneous behaviour. We jointly model GDP level and growth by further including a regression model for the variance of growth, to check for the effects of financial variables on the volatility of the growth process. Our empirical findings provide evidence of heterogeneous behaviours in both GDP level and growth rate, confirming the need of a flexible approach to properly reflect all data features. Such heterogeneous behaviours could be due to differences in institutional and technological factors and may contribute to reach (or not) economic convergence. At last, we would remark that estimated covariates effects are in line with the augmented Solow model theory, additionally the growth rate volatility is mainly related to unemployment and financial development. Of course, the model can be extended in several ways. Here, we account for heavy tails in the growth rate distribution, but other distributions than the \textit{t} one can be considered, as well as approaches to deal with outliers (if any). More than two outcomes can be jointly modelled of the price of a high computational burden involved in the estimation step. An interesting extension would deal with time-varying heterogeneity. Indeed, a limitation of our proposal is that we assume time-constant random effects. 

\newpage
\newpage
\newpage

\bibliography{references}
\bibliographystyle{te}

\newpage

\begin{table}[htbp]\centering \caption{Simulation results: Scenario 1.}\footnotesize
\begin{tabular}{l|c||ccc}\hline\hline
\multicolumn{1}{c}{\textbf{}} & \textbf{True}
 & \textbf{Estimate} & \textbf{Bias} & \textbf{Std. dev.}\\ \hline	
&&\multicolumn{3}{c}{\bf{n=100, T=5}}\\		
\hline
$u_{k_1=1}$	&	-1.00	&	-1.020	&	-0.020	&	0.265	\\
$u_{k_1=2}$	&	1.00	&	1.012	&	0.012	&	0.265	\\
$\lambda_{11}	$	&	0.50	&	0.505	&	0.005	&	0.111	\\
$u_{k_2=1}$	&	2.00	&	2.012	&	0.012	&	0.306	\\
$u_{k_2=2}$	&	-2.00	&	-2.005	&	-0.005	&	0.243	\\
$\lambda_{12}	$	&	0.50	&	0.494	&	-0.006	&	0.149	\\
\hline\hline
$\gamma_{01}$	&	0.50	&	0.489	&	-0.011	&	0.072	\\
$\gamma_{11}$	&	0.75	&	0.760	&	0.010	&	0.120	\\
$\gamma_{02}$	&	1.00	&	0.991	&	-0.009	&	0.068	\\
$\gamma_{12}$	&	0.25	&	0.253	&	0.003	&	0.122	\\
\hline\hline
$\pi_{11}$	&	0.40	&	0.420	&	0.020	&	0.048	\\
$\pi_{12}$	&	0.10	&	0.090	&	-0.010	&	0.049	\\
$\pi_{21}$	&	0.20	&	0.196	&	-0.004	&	0.047	\\
$\pi_{22}$	&	0.30	&	0.294	&	-0.006	&	0.063	\\
\hline
&\multicolumn{4}{c}{Average	Rand Index=	0.800}	\\
					\hline\hline			
&&\multicolumn{3}{c}{\bf{n=100, T=10}}\\		
\hline														
$u_{k_1=1}$	&	-1.00	&	-1.006	&	-0.006	&	0.139	\\
$u_{k_1=2}$	&	1.00	&	1.000	&	0.000	&	0.142	\\
$\lambda_{11}	$&	0.50	&	0.500	&	0.000	&	0.078	\\
$u_{k_2=1}$	&	2.00	&	2.017	&	0.017	&	0.171	\\
$u_{k_2=2}$	&	-2.00	&	-2.004	&	-0.004	&	0.134	\\
$\lambda_{12}	$	&	0.50	&	0.495	&	-0.005	&	0.098	\\
\hline\hline
$\gamma_{01}$	&	0.50	&	0.502	&	0.002	&	0.049	\\
$\gamma_{11}$	&	0.75	&	0.741	&	-0.009	&	0.086	\\
$\gamma_{02}$	&	1.00	&	0.993	&	-0.007	&	0.046	\\
$\gamma_{12}$	&	0.25	&	0.257	&	0.007	&	0.078	\\
\hline\hline
$\pi_{11}$	&	0.40	&	0.407	&	0.007	&	0.052	\\
$\pi_{12}$	&	0.10	&	0.096	&	-0.004	&	0.034	\\
$\pi_{21}$&	0.20	&	0.196	&	-0.004	&	0.045	\\
$\pi_{22}$	&	0.30	&	0.300	&	0.000	&	0.041	\\
\hline
&\multicolumn{4}{c}{Average Rand Index=	0.905}	\\
\hline
\end{tabular}\label{t:sc1}
\end{table}							

\begin{table}[htbp]\centering \caption{Simulation study: Scenario 2}\scriptsize
\begin{tabular}{l| c|| c  c c ||ccc}\hline\hline
\multicolumn{1}{c}{\textbf{}} & \textbf{True}
 & \textbf{Estimate} & \textbf{Bias} & \textbf{Std. dev.} & \textbf{Estimate} & \textbf{Bias} & \textbf{Std. dev.}\\ \hline	
&&\multicolumn{3}{c}{\bf{n=100, T=5}}&\multicolumn{3}{c}{\bf{n=100, T=10}}\\		
\hline
$u_{k_1=1}$	&	-1.00	&	-1.028	&	-0.028	&	0.337 & -1.007 & -0.007 & 0.160	\\
$u_{k_1=2}$	&	1.00	&	1.035	&	0.035	&	0.252&1.016 &0.016 & 0.123 	\\
$\lambda_{11}	$	&	0.50	&	0.498	&	-0.002	&	0.111	& 0.499 & -0.001 & 0.074\\
$u_{k_1=1}$	&	2.00	&	2.200	&	0.200	&	0.715	& 2.071 & 0.071 & 0.403\\
$u_{k_2=2}$	&	-2.00	&	-2.271	&	-0.271	&	0.935	& -2.090 & -0.090 & 0.401\\
$u_{k_2=3}$	&	0.00	&	-0.136	&	-0.136	&	0.746	&-0.066 & -0.066 & 0.616\\
$\lambda_{12}	$	&	0.50	&	0.498	&	-0.002	&	0.150	& 0.504 & 0.004 & 0.097\\
\hline\hline
$\gamma_{01}$	&	0.50	&	0.490	&	-0.010	&	0.071	&0.496 & -0.004 & 0.049\\
$\gamma_{11}$&	0.75	&	0.755	&	0.005	&	0.123	&0.751 & 0.001 & 0.084\\
$\gamma_{02}$&	1.00	&	0.989	&	-0.011	&	0.079	& 0.993& -0.007&0.050\\
$\gamma_{12}$	&	0.25	&	0.252	&	0.002	&	0.014	&0.255&0.005&0.084\\
\hline\hline
$\pi_{11}$	&	0.10	&	0.038	&	-0.062	&	0.048	&0.038 & -0.062&0.048\\
$\pi_{12}$	&	0.10	&	0.129	&	0.029	&	0.061	&0.129 & 0.029 & 0.061\\
$\pi_{13}$	&	0.20	&	0.230	&	0.030	&	0.058	&0.230 & 0.030 & 0.058\\
$\pi_{21}$	&	0.20	&	0.191	&	-0.009	&	0.075	&0.191 & -0.009 & 0.075\\
$\pi_{22}$	&	0.30	&	0.360	&	0.060	&	0.070	&0.360 &0.060&0.070\\
$\pi_{23}$	&	0.10	&	0.051	&	-0.049	&	0.058	&0.051 & -0.049 & 0.058\\
\hline
&&\multicolumn{3}{c||}{Average Rand Index=	0.740}&\multicolumn{3}{c}{Average Rand Index=	0.841}	\\
\hline\hline
\multicolumn{1}{c}{\textbf{}} & \textbf{True}
 & \textbf{Estimate} & \textbf{Bias} & \textbf{Std. dev.} & \textbf{Estimate} & \textbf{Bias} & \textbf{Std. dev.}\\ \hline	
&&\multicolumn{3}{c}{\bf{n=1000, T=5}}&\multicolumn{3}{c}{\bf{n=1000, T=10}}\\		
\hline
$u_{k_1=1}$	&	-1.00	&	-0.999&	0.001	&	0.097 & -1.000 & 0.000 & 0.050	\\
$u_{k_1=2}$	&	1.00	&	1.000	&	0.000	&	0.073	&1.002 & 0.002 & 0.039\\
$\lambda_{11}	$	&	0.50	&	0.501&	0.001	&	0.033	&0.501 & 0.001 & 0.025\\
$u_{k_2=1}$	&	2.00	&	2.054	&	0.054	&	0.215	&2.005 & 0.005 & 0.100\\
$u_{k_2=2}$	&	-2.00	&	-2.039	&	-0.039	&	0.358	& -2.005 & -0.005& 0.084\\
$u_{k_2=3}$&	0.00	&	-0.016	&	-0.016	&	0.542	&-0.002 & -0.002 & 0.185\\
$\lambda_{12}	$	&	0.50	&	0.500	&	0.000	&	0.047	&0.499 & -0.001 & 0.031\\
\hline\hline
$\gamma_{01}$	&	0.50	&	0.500	&	0.000	&	0.022	&0.500 & 0.000 & 0.015\\
$\gamma_{11}$	&	0.75	&	0.749	&	-0.001	&	0.037	& 0.751 & 0.001 & 0.026 \\
$\gamma_{02}$	&	1.00	&	1.000	&	0.000       &	0.021	& 0.999 & -0.001& 0.015\\
$\gamma_{12}$	&	0.25	&	0.249	&	-0.001	&	0.038	&0.250 & 0.000 & 0.026\\
\hline\hline
$\pi_{11}$	&	0.10	&	0.072	&	-0.028	&	0.038	&0.086 & -0.014 & 0.014\\
$\pi_{12}$	&	0.10	&	0.129	&	0.029	&	0.032	& 0.111 & 0.011 & 0.015\\
$\pi_{13}$	&	0.20	&	0.213	&	0.013	&	0.032	& 0.203 & 0.003 & 0.021\\
$\pi_{21}$	&	0.20	&	0.198	&	-0.002	&	0.041	& 0.199 & -0.001 & 0.021\\
$\pi_{22}$	&	0.30	&	0.299	&	-0.001	&	0.054	& 0.299 & -0.001 & 0.023\\
$\pi_{23}$	&	0.10	&	0.090	&	-0.010	&	0.040	& 0.101 & 0.001 & 0.026\\
\hline
&&\multicolumn{3}{c||}{Average Rand Index=	0.774}&\multicolumn{3}{c}{Average Rand Index=	0.859}	\\
\end{tabular}\label{t:sc2}
\end{table}

\begin{table}[htbp]\centering \caption{Summary statistics}
\begin{tabular}{l c c l l }\hline\hline
\multicolumn{1}{c}{\textbf{}} & \textbf{Mean}
 & \textbf{Std. Dev.} & \textbf{Variable Description} & \textbf{Sources} \\ \hline
 \textit{GDP level} &&\\
sk & 0.002 & 0.001 & share of output invested in physical capital & PWT 8.0 \\
sh & 0.632 & 0.34  & share of output invested in human capital& World Bank \\
$ng\delta$ & 0.067 & 0.012 & population growth rate $+$ $0.05^{(*)}$ & PWT 8.0\\
lnyc & 8.509 & 1.268 & log of income per capita & PWT 8.0\\
\textit{Growth} &&\\

unemp & 0.612 & 0.077  &unemployment rate & PWT 8.0\\
infl & 0.519 & 0.312 & log of consumer price & PWT 8.0\\
open & 66.7 & 38.05  & openness to trade & World Bank\\
govcons & 15.329 & 5.853  & government consumption (as share of GDP) & World Bank\\
fin & 45.656 & 39.801  & domestic credit on GDP& World Bank\\
\multicolumn{1}{c}{N} & \multicolumn{2}{c}{519}\\ \hline
\end{tabular}\label{t:sum}
\begin{footnotesize}
\begin{flushleft}

\textit{Notes}: (*): 0.05 is the commonly used value for approximating the depreciation growth rate and the technological rate. \\
\end{flushleft}
\end{footnotesize}
\end{table}

\begin{table}[ht]\centering \caption{Respone Variables: Summary statistics}\label{sumstatres} \centering
\begin{tabular}{l ccccccc}
\hline
	&	Mean	&	Std. Dev.	&	Skewness	& 	Kurtosis & Min & Max & N\\
	\hline
GDP level	&	8.6	&	1.3	&	-0.27	& 	1.96 & 5.42 & 10.70 & 519\\
GDP growth	&	0.9	&	0.2	&	-0.37	& 	10.47 &  -1.33  & 1.31 &519\\
\hline

\end{tabular}		
\end{table}

\begin{table}[ht]\centering \caption{Penalized Likelihood Criteria: Univariate model} \centering
\begin{tabular}{l lll l}
  \hline							
  \hline
	&	LLK	&		AIC	&	BIC	\\ \hline
	&		&			&		\\
$K_{1}=2$	&	-360.89	&		735.77	&	754.08	\\
$K_{1}=3$	&	-312.95	&		643.89	&	667.43	\\
$K_{1}=4$	&	-277.54	&		577.08	&	605.85	\\
$K_{1}=5$	&	-265.11	&		556.22	&	590.22	\\
$K_{1}=6$	&	-248.04	&		\textbf{526.08}	&	\textbf{565.31}	\\
$K_{1}=7$	&	-258.81	&		551.62	&	596.08	\\
 \hline
&	LLK	&		AIC	&	BIC	\\
\hline

$K_{2}=2$	&	172.22	&		\textbf{-322.44	}&	\textbf{-293.67}	\\
$K_{2}=3$	&	172.24	&		-316.47	&	-279.86	\\
$K_{2}=4$	&	173.19	&		-312.37	&	-267.91	\\
$K_{2}=5$	&	173.18	&		-306.36	&	-254.06	\\

\hline

\end{tabular}\label{t:AICBICuniv}	
\end{table}	

\begin{table}[ht]\centering \caption{Penalized Likelihood Criteria: Bivariate model}\centering
\begin{tabular}{l lll l}
  \hline							
  \hline

$K_1$	&	$K_2$	&	llk	&	AIC	&	BIC			\\ \hline
2	&	2	&	-187.32	&	414.64	&	466.94			\\
2	&	3	&	-186.55	&	421.1	&	483.86			\\
2	&	4	&	-185.51	&	427.02	&	500.24			\\
2	&	5	&	-185.57	&	435.14	&	518.82			\\
2	&	6	&	-184.91	&	441.82	&	535.96			\\
3	&	2	&	-147.21	&	340.42	&	400.57			\\
3	&	3	&	-136.14	&	328.28	&	401.50			\\
3	&	4	&	-152.45	&	370.9	&	457.20			\\
3	&	5	&	-141.97	&	359.94	&	459.31			\\
3	&	6	&	-134.52	&	355.04	&	467.49			\\
4	&	2	&	-97.18	&	246.36	&	314.35			\\
4	&	3	&	-96.29	&	256.58	&	340.26			\\
4	&	4	&	-93.43	&	262.86	&	362.23			\\
4	&	5	&	-91.77	&	271.54	&	386.61			\\
4	&	6	&	-85.44	&	270.88	&	401.64			\\
5	&	2	&	-66.97	&	191.94	&	267.78			\\
5	&	3	&	-57.64	&	187.28	&	281.42			\\
5	&	4	&	-55.25	&	196.5	&	308.95			\\
5	&	5	&	-54.27	&	208.54	&	339.30			\\
5	&	6	&	-66.52	&	247.04	&	396.10			\\
6	&	2	&	-50.42	&	164.84	&	\textbf{248.52	}		\\
6	&	3	&	-39.72	&	\textbf{159.44}	&	264.04			\\
6	&	4	&	-36.47	&	168.94	&	294.47			\\
6	&	5	&	-61.48	&	234.96	&	381.41			\\
6	&	6	&	-35.29	&	198.58	&	365.95			\\
\hline

\end{tabular}\label{t:AICBICbiv} 

\end{table}	

\begin{table}[ht]\centering \caption{Results} \centering\tiny
\begin{tabular}{l llr ll lr}
  \hline							
  \hline & &\multicolumn{3}{c}{ \textbf{Univariate} } & \multicolumn{3}{c}{ \textbf{Bivariate} } \\ \hline
   												
        	&	&	Coef.	&	SE &	&	&	Coef	&	SE \\ \hline
        	{\large \textit{Income per capita:}}\\
\textit{$\mu_{it1}$ }\\      	
 \textit{sk}    	&	&	      0.07    **      	&	0.03	&	&	&	0.14    ***	&	0.03	\\
 \textit{sh}    	&	&	      0.72    ***     	&	0.02	&	&	&	0.46    ***	&	0.03	\\
  $ng\delta$  	&	&	       -0.33   ***     	&	0.10	&	&	&	-0.61   ***	&	0.1	\\
$u_{0k_{1}=1}    $	&	&	      8.46    ***     	&	0.32	&	&	&	9.64    ***	&	0.31	\\
$u_{0k_{1}=2}    $	&	&	      9.02    ***     	&	0.06	&	&	&	7.48    ***	&	0.31	\\
$u_{0k_{1}=3}    $	&	&	      9.59    ***     	&	0.06	&	&	&	8.07    ***	&	0.3	\\
$u_{0k_{1}=4}$   	&	&	       10.19   ***     	&	0.06	&	&	&	6.97    ***	&	0.31	\\
$u_{0k_{1}=5}$   	&	&	       10.84   ***     	&	0.06	&	&	&	8.59    ***	&	0.3	\\
$u_{0k_{1}=6}    $	&	&	      11.29   ***     	&	0.09	&	&	&	9.01    ***	&	0.3	\\
  \\
  
  $log(\sigma_{it1})$ \\                                                    										
$\gamma_{01}$   	&	&	       -1.23   ***     	&	0.03	&	&	&	-1.28   ***	&	0.03	\\
\\

Observations &  &\multicolumn{3}{l}{ 519} & \multicolumn{3}{l}{ 519 }\\
$K_{1}$ &  &\multicolumn{3}{l}{ 6} & \multicolumn{3}{l}{ 6}\\
$\ell^{(*)}$ &  &\multicolumn{3}{l}{ -248.81} & \multicolumn{3}{l}{ }\\
$\ell$ &  &\multicolumn{3}{l}{ } & \multicolumn{3}{l}{ -50.42}\\
\hline
	{\large \textit{Growth rate:}}\\
	
 $\mu_{it2}$\\ 											
$u_{0k_{2}=1}$   	&	&	-0.01	&	0.05	&	&	&	1.05    ***	&	0.12	\\
$u_{0k_{2}=2}$   	&	&	       1.11    ***     	&	0.15	&	&	&	-0.1	&	0.09	\\
  $lnyc _{k_{2}=1}$       	&	&	       0.01    **      	&	0.01	&	&	&	-0.09   ***	&	0.12	\\
  $lnyc_{k_{2}=2}$        	&	&	       -0.1    ***     	&	0.02	&	&	&	0.02    **	&	0.01	\\
  \\
  
  $log(\sigma_{it2})$\\                                                    												
$\gamma_{02}$	&	&	  -1.53   ***     	&	0.52	&	&	&	-1.52   ***	&	0.51	\\
 \textit{unemp  	}&	&	       1.17    **      	&	0.55	&	&	&	1.34    **	&	0.53	\\
  \textit{infl  	}&	&	0.01	&	0.08	&	&	&	0.03	&	0.08	\\
\textit{open    	}&	&	0.05	&	0.08	&	&	&	0.05	&	0.08	\\
\textit{govcons }	&	&	-0.11	&	0.11	&	&	&	-0.15	&	0.11	\\
\textit{fin }    	&	&	       -0.31   ***     	&	0.05	&	&	&	-0.31   ***	&	0.05	\\
 \\

  $\nu_{it2}$\\                                                    												
$\tilde{\gamma}$        	&	&	       1.72    ***     	&	0.17	&	&	&	1.69    ***	&	2.23	\\
\\

Observations &  &\multicolumn{3}{l}{ 519} & \multicolumn{3}{l}{ 519 }\\
$k_{2}$ &  &\multicolumn{3}{l}{ 2} & \multicolumn{3}{l}{ 2}\\
$\ell^{(*)}$ &  &\multicolumn{3}{l}{ 172.22} & \multicolumn{3}{l}{ }\\
$\ell$ &  &\multicolumn{3}{l}{ } & \multicolumn{3}{l}{ -50.42}\\
					\hline\hline
\end{tabular}\label{res}

\begin{footnotesize}
\begin{flushleft}
{Significance level}:\hspace{1em} $***$ : 0.1\%  \hspace{1em} $**$ : 1\%
\hspace{1em} $*$ : 5\%\\
\emph{Notes}:  $\ell^{(*)}$: log-likelihood for the univariate model, $\ell$: log-likelihood for the bivariate model. Dependent variables: 5 years forward value of log of GDP per capita (top of the Table), and 5 years forward value of growth rate. 
\end{flushleft}

\end{footnotesize}				
\end{table}		

\begin{table}[ht]\centering \caption{Clustering results}\centering\label{group}\scriptsize
\begin{tabular}{l| l|l|}
& \multicolumn{2}{c}{$K_2$}\\
\hline
$K_1$ & \multicolumn{1}{c}{$1$} & \multicolumn{1}{c}{$2$} \\
\hline
1 & Australia, Austria, Belgium,& \\&Canada, Czech Rep., Denmark,&\\ & Finland, France,
 Germany,&\\ & Hong Kong, Ireland, Israel,&\\ & Italy, Japan, Netherlands,&\\ & New Zealand, Norway, Spain,&
\\ & Sweden, Switzerland, Trinidad \& Tobago, UK, USA &  \\
\hline
2 & & Bangladesh, Benin, Burkina Faso, \\ && Burundi, Rep. Congo, India, \\ && Kenya, Madagascar, Mali, \\ && Moldova, Niger, Rwanda,\\ && Sri Lanka, Syria, Tanzania, Uganda\\
\hline
3 & China & Bolivia, Cameroon, Chad,\\ && Djibouti, Egypt, Honduras, \\ && Indonesia, Jamaica, Jordan, \\ &&Mauritania, Morocco, Pakistan,\\ && Paraguay, Peru, Phillippines,\\ && Senegal, Sierra Leone\\\hline
4 & &Rep. Central African, Rep. Dem. Congo,\\&& El Salvador, Malawi, Mozambique,\\&& Nepal, Nigeria, Togo\\\hline
5 & Thailand & Bulgaria, Colombia, Dominican Rep.,\\ && Ecuador, Guatemala, Serbia,\\ && South Africa, Tunisia, Uruguay,Zimbabwe\\ \hline
6 & Angola, Argentina, Botswana, & Costa Rica, Mexico, Panama, Turkey, Venezuela\\ & Chile, Croatia, Estonia, Greece,&\\& Hungary, Rep. of Korea, Latvia,&\\ & Malaysia, Maldives, Mauritius,&\\& Poland, Poland, Portugal,&\\ & Romania, Russia, Slovakia& \\
\hline	

\end{tabular}\label{t:clust}
\end{table}

\newpage
\newpage

\end{document}